# Protocol for Executing and Benchmarking Eight Computational Doublet-Detection Methods in Single-Cell RNA Sequencing Data Analysis

**Nan Miles Xi [1] and Jingyi Jessica Li [1,2,3,4,5,]***

[1]Department of Statistics, University of California, Los Angeles, CA 90095-1554, USA
[2]Department of Human Genetics, University of California, Los Angeles, CA 90095-7088, USA
[3]Department of Computational Medicine, University of California, Los Angeles, CA 90095-1766, USA
[4]Department of Biostatistics, University of California, Los Angeles, CA 90095-1772, USA
[5]Lead Contact
*Correspondence: jli@stat.ucla.edu

## Summary

The existence of doublets is a key confounder in single-cell RNA sequencing (scRNA-seq) data analysis. Computational methods have been developed for detecting doublets from scRNA-seq data. We develop an R package DoubletCollection to integrate the installation and execution of eight doublet-detection methods. DoubletCollection also provides a unified interface to perform and visualize downstream analysis after doublet detection. Here, we present a protocol of using DoubletCollection to benchmark doublet-detection methods. This protocol can automatically accommodate new doublet-detection methods in the fast-growing scRNA-seq field.

For detailed results from the use of this protocol, please refer to (N. M. Xi and Li 2020).

 **CellPress**

## Graphical Abstract

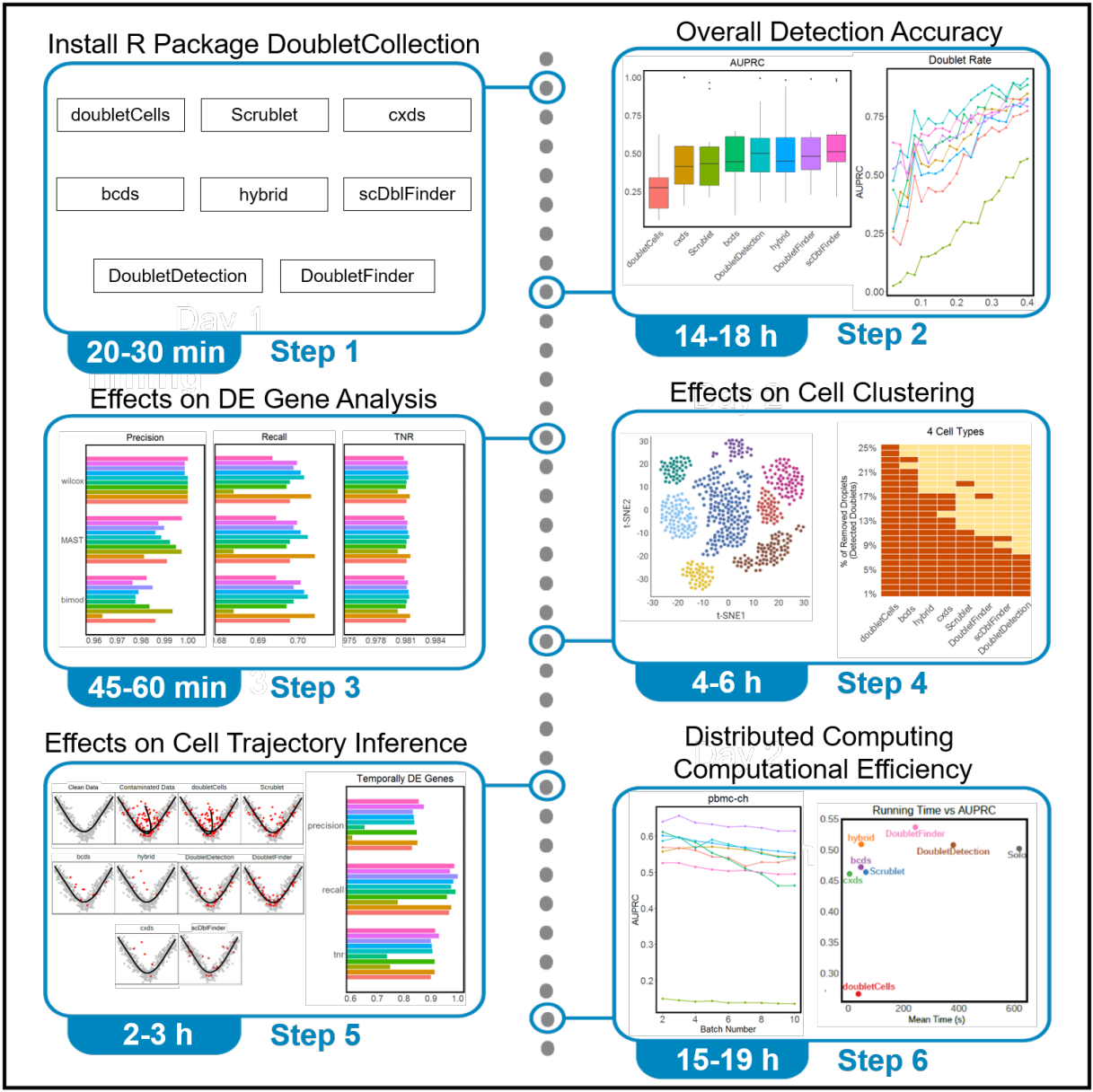

Install R Package DoubletCollection
- doubletCells | Scrublet | cxds
- bcds | hybrid | scDblFinder
- DoubletDetection | DoubletFinder

**20-30 min** **Step 1**

Overall Detection Accuracy

**14-18 h** **Step 2**

Effects on DE Gene Analysis

**45-60 min** **Step 3**

Effects on Cell Clustering

**4-6 h** **Step 4**

Effects on Cell Trajectory Inference

**2-3 h** **Step 5**

Distributed Computing
Computational Efficiency

**15-19 h** **Step 6**

## Before You Begin

This section describes how to download real and synthetic scRNA-seq datasets used in this protocol from a public web repository. We also introduce the installation of DoubletCollection. For consistency, we will refer to a "cell" (either a singlet or a doublet/multiplet) in a scRNA-seq dataset as a "droplet" in the following text.

## Download real scRNA-seq datasets with experimentally annotated doublets

**Timing: 10-20 min**

We collect 16 real scRNA-seq datasets with doublets annotated by experimental techniques. This collection covers a variety of cell types, droplet and gene numbers, doublet rates, and sequencing depths. It represents varying levels of difficulty in detecting doublets from scRNA-seq data. The data collection and preprocessing details are described in (N. M. Xi and Li 2020). The datasets are available at Zenodo https://zenodo.org/record/4562782#.YI2lhWf0mbg in the file `real_datasets.zip` (Figure 1).

The datasets are in the `rds` format. The name of each dataset file is the same as the name defined in (N. M. Xi and Li 2020). After being loaded into R, each dataset is a list containing two elements: the first element is a scRNA-seq count matrix with rows as genes and columns as droplets; the second element is a vector containing the singlet/doublet annotation of each droplet, which corresponds to each column in the first element.

## Download synthetic scRNA-seq datasets under various experimental settings and biological conditions

**Timing: 10-20 min**

We utilize two simulators, scDesign (Li and Li 2019) and Splatter (Zappia, Phipson, and Oshlack 2017), to generate realistic scRNA-seq datasets with varying doublet rates (i.e., percentages of doublets among all droplets), sequencing depths, cell types, and between-cell-type heterogeneity levels. The synthetic datasets contain ground-truth doublets, cell types, differentially expressed (DE) genes, and cell trajectories. The simulation details are described in (N. M. Xi and Li 2020). The datasets are available at Zenodo https://zenodo.org/record/4562782#.YI2lhWf0mbg in the file `synthetic_datasets.zip` (Figure 1). The datasets are in the `rds` format. All the synthetic datasets contain count matrices with rows as genes and columns as droplets. Below is the data structure of each dataset after being loaded into R.

**sim_rate.rds:** An R list with two elements. The first element contains 20 scRNA-seq count matrices that are independently generated with the doublet rates ranging from 0.02 to 0.4. Each element is named by its doublet rate and has rows as genes and columns as droplets. The second element contains 20 singlet/doublet annotation vectors, which correspond to the columns of the 20 count matrices in the first element.

**sim_depth.rds:** An R list with two elements. The first element contains 20 scRNA-seq count matrices that are independently generated with sequencing depth ranging from 500 to 10,000 UMI counts. Each element is named by its sequencing depth and has rows as genes and columns as droplets. The second element contains 20 singlet/doublet annotation vectors, which correspond to the columns of the 20 count matrices in the first element.

**sim_type.rds:** An R list with two elements. The first element contains 19 scRNA-seq count matrices that are independently generated with cell type numbers ranging from 2 to 20. Each element is named by its cell type numbers and has rows as genes and columns as droplets. The second element contains 19 singlet/doublet annotation vectors, which correspond to the columns of the 19 count matrices in the first element.

**sim_hetero.rds:** An R list with two elements. The first element contains 21 scRNA-seq count matrices that are independently generated with different between-cell-type heterogeneity levels as defined in (N. M. Xi and Li 2020). Each element is named by its between-cell-type heterogeneity levels and has rows as genes and columns as droplets. The second element contains 21 singlet/doublet annotation vectors, which correspond to the columns of the 21 count matrices in the first element.

**sim_clustering.rds:** An R list with two elements. The first element contains three scRNA-seq count matrices with four, six, or eight cell types. Each element is named by its cell type numbers and has rows as genes and columns as droplets. The second element contains three singlet/doublet annotation vectors, which correspond to the columns of the three count matrices in the first element.

**sim_DE.rds:** An R list with four elements, including one synthetic scRNA-seq count matrix, its doublet indices, cell type annotations, and DE genes. This dataset contains 6% DE genes between two cell types and 40% doublets.

**sim_trajectory.rds:** An R list with two elements, including one synthetic scRNA-seq count matrix and its singlet/doublet annotations. This dataset contains a bifurcating cell trajectory and 20% doublets.

**sim_temporally_DE.rds:** An R list with three elements, including one synthetic scRNA-seq count matrix, its singlet/doublet annotations, and temporally DE genes. This dataset contains one cell trajectory with 250 temporally DE genes and 20% doublets.

**Note**: Please download the latest version of the datasets from the Zenodo repository. The timing for data downloading depends on the network condition.

## Key Resources Table

| REAGENT or RESOURCE | SOURCE | IDENTIFIER |
|---|---|---|
| Deposited Data | | |
| 16 real scRNA-seq datasets with doublet annotations | (N. M. Xi and Li 2020) | https://zenodo.org/record/4562782#.YI2lhWf0mbg |
| Synthetic scRNA-seq datasets with ground-truth doublets, cell types, DE genes, and cell trajectories | (N. M. Xi and Li 2020) | https://zenodo.org/record/4562782#.YI2lhWf0mbg |
| Software and Algorithms | | |
| R | https://www.r-project.org/ | https://www.r-project.org/ |

| RStudio | https://www.rstudio.com/products/rstudio/download/ | https://www.rstudio.com/products/rstudio/download/ |
| Python 3 | https://www.anaconda.com/products/individual | https://www.anaconda.com/products/individual |
| devtools | https://cran.r-project.org/web/packages/devtools/devtools.pdf | https://cran.r-project.org/web/packages/devtools/index.html |
| Rtools | https://cran.r-project.org/bin/windows/Rtools/ | https://cran.r-project.org/bin/windows/Rtools/ |
| DoubletCollection | This paper | https://github.com/xnnba1984/DoubletCollection |

# Materials and Equipment

- **Hardware**
  - We recommend 64 GB or more memory due to the large sizes of real and synthetic datasets in this protocol. The required memory would be less if users perform doublet detection on smaller datasets.
  - We recommend a CPU with four or more cores to conduct the benchmark study.
- **Software**
  The installation of DoubletCollection will automatically install all required R packages. The Python 3 environment is required since two doublet-detection methods included in DoubletCollection are developed in Python 3. We recommend installing Anaconda for the Python 3 environment.
  - R (v 4.0.5)
  - Python 3 (v 3.8.5)
  - RStudio (v 1.4.1)
  - devtools (v 2.4.0)
  - Rtools (v 4.0)
  - DoubletCollection (v 1.1.0)

# Step-by-Step Method Details

## Install DoubletCollection

**Timing: 20-30 min**

DoubletCollection is an R package that integrates the installation, execution, and benchmark of eight doublet-detection methods. The source code and documentation of DoubletDetection are available at

[https://github.com/xnnba1984/DoubletCollection](https://github.com/xnnba1984/DoubletCollection). To install DoubletDetection, execute the following R code.

```
if(!require(devtools)){
  install.packages("devtools")
}
devtools::install_github("xnnba1984/DoubletCollection")
```

**Note**: DoubletCollection automatically installs eight doublet-detection methods: Scrublet (Wolock, Lopez, and Klein 2019), doubletCells (Lun, McCarthy, and Marioni 2016), scds (Bais and Kostka 2019) (including cxds, bcds, and hybrid), DoubletDetection (Gayoso and Shor 2019), DoubletFinder (McGinnis, Murrow, and Gartner 2019), and scDblFinder (Germain, Sonrel, and Robinson 2020). It also installs other packages required for downstream analysis and visualization.

**Optional**: Solo (Bernstein et al. 2020) is a doublet-detection method implemented as a Linux command-line tool. DoubletCollection does not include this method. The installation and execution of Solo are available at [https://github.com/calico/solo](https://github.com/calico/solo).

## Doublet detection accuracy on real scRNA-seq datasets

**Timing: 6-8 h**

This section illustrates how to apply DoubletCollection to 16 real scRNA-seq datasets, calculate the detection accuracy, and visualize the result.

1. **Calculate doublet scores**

   Every doublet-detection method in DoubletCollection outputs a doublet score for each droplet in the dataset. The larger the doublet score is, the more likely the droplet is a doublet. The following R code calculates doublet scores of user-specified methods on 16 real datasets.

   ```
   library(DoubletCollection)

   # read 16 datasets in the directory specified by the path parameter
   data.list <- ReadData(path = "directory/real_datasets") # Do not directly
   copy this line ("directory" is where "real_datasets.zip" is downloaded
   and uncompressed)
   count.list <- data.list$count

   # transform doublet annotations to 0/1
   label.list <- lapply(data.list$label, FUN = function(label){
     ifelse(label == 'doublet', 1, 0)
   })
   methods <- c('doubletCells','cxds','bcds','hybrid','scDblFinder',
                'Scrublet','DoubletDetection','DoubletFinder')

   # calculate doublet scores
   score.list.all <- FindScores.All(count.list, methods)
   ```

**Note**: In function `ReadData`, users need to set the `path` parameter to the directory where they save datasets. Suppose that the local directory of 16 real datasets is "F:\Dropbox\doublet\package\real_datasets" on a Windows machine (Figure 6A). To read these datasets, users need to execute

```
data.list <- ReadData(path = "F:/Dropbox/doublet/package/real_datasets")
```

(If users need to find the directory path, they may drag the directory folder to the R console, and the path will show up.) After execution, `data.list` will be a list that contains 16 datasets (Figure 6B). Users do not necessarily need to download all 16 datasets to run the test. Instead, users can save a subset of those datasets in a directory and just set the `path` parameter to the path of that directory. Users can also choose doublet-detection methods by modifying the `methods` vector.

If `methods` includes `'DoubletFinder'`, a figure (about hyperparameter search) would be automatically generated by the DoubletFinder method.

2. **Calculate the area under the precision-recall curve (AUPRC) and the area under the receiver operating characteristic curve (AUROC)**
   Doublet detection is essentially a binary classification problem. Therefore, AUPRC and AUROC are appropriate for evaluating the overall doublet-detection accuracy. The following R code calculates AUPRC and AUROC based on the doublet scores.

```
auprc.list.all <- FindAUC.All(score.list.all, label.list, 'AUPRC')
auroc.list.all <- FindAUC.All(score.list.all, label.list, 'AUROC')
```

3. **Visualize overall doublet-detection accuracy**
   We use boxplots to visualize the distributions of AUPRC and AUROC values of every doublet-detection method on the 16 real scRNA-seq datasets. The following R code outputs Figure 2A.

```
# transform the output of FindAUC.All to a data frame for visualization
result.auprc <- ListToDataframe(auprc.list.all, 'boxplot')
result.auroc <- ListToDataframe(auroc.list.all, 'boxplot')
```

**Note**: Users can save data frames `result.auprc` and `result.auroc` to compare the AUPRC and AUROC values of doublet-detection methods.

```
# visualize AUPRC and AUROC by boxplots
# save each plot to a file in the current working directory
Plot_Boxplot(result.auprc, 'AUPRC', save=T, name = 'AUPRC_real.png', path
= getwd())
Plot_Boxplot(result.auroc, 'AUROC', save=T, name = 'AUROC_real.png', path
= getwd())
```

**Note**: Users can save each plot to the local directory by setting the `save` parameter to "T."

4. **Calculate precision, recall, and true negative rate (TNR) under specific identification rates**
   In practice, doublets are identified based on a single threshold. To accommodate this scenario, we examine the detection accuracy of doublet-detection methods under a specific identification rate x%. For each method and each dataset, we identify the top x% droplets with the highest doublet scores as doublets. Then we calculate the corresponding precision, recall, and TNR. The following R code calculates precision, recall, and TNR under a 10% identification rate.

```
# call doublets based on a 10% doublet rate
doublet.list.all <- FindDoublets.All(score.list.all, rate=0.1)

# calculate precision, recall, and TNR of identified doublets
precision.list.all <- FindACC.All(doublet.list.all, label.list,
                                  'precision')
recall.list.all <- FindACC.All(doublet.list.all, label.list, 'recall')
tnr.list.all <- FindACC.All(doublet.list.all, label.list, 'TNR')
```

**Optional**: Users can calculate the precision, recall, and TNR under varying doublet rates to conduct a more comprehensive comparison of doublet-detection methods.

5. **Visualize precision, recall, and TNR under specific identification rates**

   Again, we use boxplots to visualize the distributions of precision, recall, and TNR values of each method under specific identification rates. The following R code outputs Figure 2B.

```
# transform the output of FindACC.All to a data frame for visualization
result.precision <- ListToDataframe(precision.list.all, 'boxplot')
result.recall <- ListToDataframe(recall.list.all, 'boxplot')
result.tnr <- ListToDataframe(tnr.list.all, 'boxplot')
```

**Note**: Users can save data frames `result.precision`, `result.recall`, and `result.tnr` to compare the precision, recall, and TNR values of doublet-detection methods.

```
# visualize precision, recall, and TNR by boxplots
# save each plot to a file in the current working directory
Plot_Boxplot(result.precision, 'Precision', save=T,
             name = 'precision.png', path = getwd())
Plot_Boxplot(result.recall, 'Recall', save=T,
             name = 'recall.png', path = getwd())
Plot_Boxplot(result.tnr, 'TNR', save=T, name = 'TNR.png', path = getwd())
```

**Note**: Users can save each plot to the local directory by setting the `save` parameter to "T."

## Hyperparameter tuning for doublet-detection methods (Optional)

**Timing: 3-4 h**

The previous R code sets the hyperparameters of doublet-detection methods to their recommended or default values. This section explains how to use DoubletCollection to search for the hyperparameters that may potentially improve the doublet-detection methods' performance.

1. **Search for optimal hyperparameters**

   We set up a series of hyperparameter values and use DoubletCollection to conduct a grid search. DoubletCollection returns a combination of hyperparameters that optimizes a user-specified accuracy measure on a dataset. The following R code searches for optimal hyperparameters in terms of AUPRC for the methods Scrublet, DoubletFinder, and scDblFinder on the dataset `pbmc-1A-dm`.

```
# read dataset
count <- count.list$`pbmc-1A-dm`
label <- label.list$`pbmc-1A-dm`
set.seed(2021)

# search for optimal hyperparameters of Scrublet
result.parameter.Scrublet <- FindParameters(count, label,
                                method = 'Scrublet', type = 'AUPRC',
                                n_neighbors = c(29, 30, 31),
                                n_prin_comps = c(30, 35, 40),
                  min_gene_variability_pctl = c(60, 65, 70))
# print optimal hyperparameters
result.parameter.Scrublet

# search for optimal hyperparameters of DoubletFinder
result.parameter.DoubletFinder <- FindParameters(count, label,
                                method = 'DoubletFinder', type = 'AUPRC',
                                nfeatures = c(1000, 1500, 2000),
                                PCs = c(10, 15, 20))
# print optimal hyperparameters
result.parameter.DoubletFinder

# search for optimal hyperparameters of scDblFinder
result.parameter.scDblFinder <- FindParameters(count, label,
                                method = 'scDblFinder', type = 'AUPRC',
                                nf=c(500, 1000, 1500),
                                includePCs=c(5, 6, 7),
                                max_depth=c(4, 5, 6))
# print optimal hyperparameters
result.parameter.scDblFinder
```

**Note**: The optimal hyperparameter found by scDblFinder is stored in `result.parameter.scDblFinder`. For the purpose of testing the code, users may reduce the number of candidate values for each parameter. Users can search for optimal hyperparameters of other doublet-detection methods on any datasets. The searchable hyperparameters of a doublet-detection method can be shown by executing `?FindParameters`.

2. **Adjust hyperparameters for doublet-detection methods**
The optimal hyperparameters found from a representative dataset provide guidance for applying a doublet-detection method to similar datasets. The following R code sets the hyperparameters of three doublet-detection methods, which are to be applied to the dataset `pbmc-1B-dm`, to their optimal values found from the dataset `pbmc-1A-dm`. These two datasets share the same cell types and experimental protocol.

```
score.list <- FindScores(count = count.list$`pbmc-1B-dm`,
        methods = c('Scrublet','DoubletFinder','scDblFinder'),
        n_neighbors=29, min_gene_variability_pctl=60, n_prin_comps=40,
        nfeatures=1000, PCs=10,
        nf=500, includePCs=6, max_depth=4)
```

**Note**: All adjustable hyperparameters can be found by executing `?FindScores` or `?FindScores.All`. Users can also adjust hyperparameters based on their prior knowledge. The R code in the following sections uses the recommended or default hyperparameter values of doublet-detection methods.

## Doublet detection accuracy on synthetic scRNA-seq datasets under various experimental settings and biological conditions

**Timing: 8–10 h**

This section illustrates how to apply DoubletCollection to synthetic scRNA-seq datasets under a wide range of experimental settings and biological conditions, calculate the detection accuracy of different doublet-detection methods, and visualize the result.

1. **Calculate doublet scores**

   As in the previous section, we first calculate doublet scores on synthetic datasets. The following R code calculates doublet scores on the dataset `sim_rate` with different doublet rates.

   ```
   # specify your directory of dataset sim_rate
   data.list.sim <- readRDS("directory/synthetic_datasets/sim_rate.rds")
   # Do not directly copy this line ("directory" is where
   "synthetic_datasets.zip" is downloaded and uncompressed)
   count.list.sim <- data.list.sim$count
   label.list.sim <- lapply(data.list.sim$label, FUN = function(label){
     ifelse(label == 'doublet', 1, 0)
   })
   score.list.all.sim <- FindScores.All(count.list.sim, methods)
   ```

   **Note**: In function `readRDS`, users need to set the directory to the parent folder of `synthetic_datasets`. Users can read datasets `sim_depth.rds`, `sim_type.rds`, or `sim_hetero.rds` to calculate doublet scores under various sequencing depth, number of cell types, or degree of between-cell-type heterogeneity. The code in the following sections can be applied to those datasets without modification.

2. **Calculate AUPRC and AUROC**

   Similar to real datasets, we use AUPRC and AUROC to measure the overall detection accuracy on synthetic datasets. The following R code calculates AUPRC and AUROC based on the doublet scores obtained from the previous step.

   ```
   auprc.list.all.sim <- FindAUC.All(score.list.all.sim, label.list.sim,
                                     'AUPRC')
   auroc.list.all.sim <- FindAUC.All(score.list.all.sim, label.list.sim,
                                     'AUROC')
   ```

3. **Visualize doublet-detection accuracy under various experimental settings and biological conditions**

We use line plots to show how the performance of each doublet-detection method changes when we vary the experimental settings and biological conditions. The following R code draws line plots for AUPRC and AUROC under varying doublet rates. Figure 3A shows the AUPRC and AUROC values of different doublet-detection methods under different doublet rates, sequencing depths, numbers of cell types, and heterogeneity between cell types.

```
# transform the output of FindAUC.All to a data frame for visualization
result.auprc.sim <- ListToDataframe(auprc.list.all.sim, 'lineplot')
result.auroc.sim <- ListToDataframe(auroc.list.all.sim, 'lineplot')

# visualize AUPRC and AUROC by line plots
# save each plot to a file in the current working directory
Plot_Lineplot(result.auprc.sim, 'Doublet Rate', 'AUPRC', save=T,
              name = 'auprc_rate.png', path = getwd())
Plot_Lineplot(result.auroc.sim, 'Doublet Rate', 'AUROC', save=T,
              name = 'auroc_rate.png', path = getwd())
```

## Effects of doublet detection on DE gene analysis

**Timing: 45-60 min**

This section illustrates how to use DoubletCollection to conduct differentially expressed (DE) gene analysis. We compare the results of DE gene analysis on the contaminated dataset (with 40% doublets), the clean dataset (without doublets), and the dataset after each doublet-detection method is applied.

1. **Detect and remove doublets from datasets**
   We first read in the dataset `sim_DE` that includes the ground-truth DE genes and 40% doublets. Then we apply doublet-detection methods to obtain doublet scores. Finally, we remove the top 40% droplets that receive the highest doublet scores from each method.

```
# specify your directory of dataset sim_DE
data.de <- readRDS('directory/synthetic_datasets/sim_DE.rds') # Do not
directly copy this line ("directory" is where "synthetic_datasets.zip" is
downloaded and uncompressed)
score.list.de <- FindScores(data.de$count, methods)
doublet.list.de <- FindDoublets(score.list.de, rate=0.4)

# add the clean data matrix to the data list
doublet.list.de[['Clean Data']] <- data.de$label.doublet

# remove identified doublets
data.removal.list.de <- RemoveDoublets.Method(data.de$count,
                data.de$label.cluster, doublet.list.de)

# add original contaminated data to the data list
data.removal.list.de[['Contaminated Data']] <- list(count=data.de$count,
                label=data.de$label.cluster)
```

2. **Identify DE genes**

We use the Wilcoxon rank-sum test (Fay and Proschan 2010), MAST (Finak et al. 2015), and likelihood-ratio test (bimod) (McDavid et al. 2013) to identify DE genes between two cell types. The accuracy of DE gene identification is measured by precision, recall, and TNR.

```
# create a data frame to save result for visualization
table.DE.all <- data.frame()

# use three DE methods
for(DE.method in c('MAST', 'wilcox', 'bimod')){

    # identify DE genes
    DE.list <- FindDE(data.removal.list.de, DE.method)

    # calculate precision, recall, and TNR of identified DE genes
    DE.acc.list <- FindDEACC(DE.list, data.de$gene.de,
                    rownames(data.de$count))

    # transform to a data frame for visualization
    table.DE <- ListToDataframe(DE.acc.list, 'barplot')
    table.DE[['DE_method']] <- DE.method
    table.DE.all <- rbind(table.DE.all, table.DE)
}
```

**Note**: Users can choose from seven DE methods by specifying the second parameter of the function `FindDE`, including 'wilcox', 'bimod', 't', 'poisson', 'negbinom', 'LR', and 'MAST'. A detailed demonstration of those methods is available at
https://satijalab.org/seurat/articles/de_vignette.html.

3. **Visualize the effects of doublet detection on DE gene analysis**
   We use barplots to compare the results of DE gene analysis on the contaminated dataset (negative control), the clean dataset (positive control), and post-doublet-detection datasets. The following R code outputs barplots that compare the precision, recall, and TNR in Figure 3B. Each barplot stacks the results of three DE methods: Wilcoxon rank-sum test, MAST, and likelihood-ratio test (bimod).

```
# save each plot to a file in the current working directory
Plot_Barplot(table.DE.all[table.DE.all$measurement=='precision',],
        'Precision', save=T, name = 'precision_DE.png', path = getwd())
Plot_Barplot(table.DE.all[table.DE.all$measurement=='recall',],
        'Recall', save=T, name = 'recall_DE.png', path = getwd())
Plot_Barplot(table.DE.all[table.DE.all$measurement=='tnr',],
        'TNR', save=T, name = 'TNR_DE.png', path = getwd())
```

## Effects of doublet detection on cell clustering

**Timing: 4-6 h**

This section illustrates how to use DoubletCollection to evaluate the effects of doublet-detection methods on cell clustering. First, we examine the efficacy of doublet-detection methods for removing spurious cell clusters formed by doublets. Second, we compare the proportion of singlets in the correctly identified cell clusters after each doublet-detection method is applied.

1. **Detect and remove doublets from datasets**

   We first read the dataset `sim_clustering` that includes three datasets with four, six, and eight cell types and 20% doublets. Then we apply doublet-detection methods to obtain doublet scores and remove doublets based on various doublet rates.

   ```
   # specify your directory of dataset sim_clustering
   data.list.cluster <-
    readRDS("directory/synthetic_datasets/sim_clustering.rds")  # Do not
   directly copy this line ("directory" is where "synthetic_datasets.zip" is
   downloaded and uncompressed
   count.list.cluster <- data.list.cluster$count
   label.list.cluster <- lapply(data.list.cluster$label,
                       FUN = function(label){
                          ifelse(label == 'doublet', 1, 0)
                       })
   score.list.all.cluster <- FindScores.All(count.list.cluster, methods)

   # call doublets based on doublet rates from 0.01 to 0.25
   doublet.list.all.rate.cluster <-
                        FindDoublets.All.Rate(score.list.all.cluster,
                             rates = seq(0.01, 0.25, 0.01))

   # remove identified doublets under different doublet rates
   data.removal.all.rate.cluster <-
          RemoveDoublets.All.Rate(count.list.cluster, label.list.cluster,
                                doublet.list.all.rate.cluster)
   ```

2. **Perform cell clustering after doublet detection**

   We apply Louvain clustering (Blondel et al. 2008) to the post-doublet-removal datasets to identify cell clusters.

   ```
   result.cluster.all.rate <-
                        Clustering.All.Rate(data.removal.all.rate.cluster)
   ```

3. **Visualize the effects of doublet detection on the removal of spurious cell clusters**

   We use heatmaps to compare the efficacy of doublet-detection methods for removing spurious cell clusters. The following R code outputs heatmaps of clustering results on datasets with four, six, and eight cell clusters under various doublet rates (Figure 4A).

   ```
   # transform the output of Clustering.All.Rate to a data frame for
   # visualization
   table.cluster <- ListToDataframe(result.cluster.all.rate, type='heatmap')

   # draw heatmaps of clustering results
   # save each plot to a file in the current working directory
   Plot_Heatmap(table.cluster, cluster = 4, save=T,
                name = 'clustering_4.png', path = getwd())
   Plot_Heatmap(table.cluster, cluster = 6, save=T,
                name = 'clustering_6.png', path = getwd())
   ```

```
Plot_Heatmap(table.cluster, cluster = 8, save=T,
             name = 'clustering_8.png', path = getwd())
```

**4. Calculate singlet proportions after doublet detection**

Homotypic doublets tend to cluster together with singlets and thus do not form spurious clusters. To evaluate the efficacy of doublet-detection methods for eliminating homotypic doublets, we calculate the proportion of singlets in each identified cell cluster when the number of cell clusters matches the number of cell types.

```
table.cluster.quality <- Clustering.Quality(table.cluster,
                              result.cluster.all.rate,
                              data.removal.all.rate.cluster)
```

**5. Visualize the singlet proportions in cell clusters**

We use boxplots to visualize the singlet proportions within clusters after applying doublet-detection methods, if the remaining droplets lead to the correct number of cell clusters (Figure 4B).

```
# save each plot to a file in the current working directory
Plot_Boxplot(table.cluster.quality[table.cluster.quality$correct=='4',],
             'Singlet Rates (Four Clusters)', save=T,
             name = 'cluster_quality_4.png', path = getwd())
Plot_Boxplot(table.cluster.quality[table.cluster.quality$correct=='6',],
             'Singlet Rates (Six Clusters)', save=T,
             name = 'cluster_quality_6.png', path = getwd())
Plot_Boxplot(table.cluster.quality[table.cluster.quality$correct=='8',],
             'Singlet Rates (Eight Clusters)', save=T,
             name = 'cluster_quality_8.png', path = getwd())
```

## Effects of doublet detection on cell trajectory inference

**Timing: 2-3 h**

This section illustrates how to use DoubletCollection to evaluate the effects of doublet-detection methods on cell trajectory inference. First, we examine the efficacy of doublet-detection methods for removing spurious cell branches formed by doublets. Second, we compare the accuracy of temporally DE gene identification after doublet-detection methods are applied.

**1. Perform and visualize cell trajectory inference on the contaminated dataset**

We use Slingshot (Street et al. 2018) to infer the cell trajectories on the dataset `sim_trajectory`. It contains two cell branches mixed with 20% doublets (contaminated dataset). The following R code shows a two-dimensional visualization of the inference result. It contains three cell trajectories instead of two, and the intermediate trajectory is formed by doublets (Figure 5A).

```
# specify your directory of dataset sim_trajectory
data.trajectory <-
  readRDS('directory/synthetic_datasets/sim_trajectory.rds')  # Do not
directly copy this line ("directory" is where "synthetic_datasets.zip" is
downloaded and uncompressed
```

```
count <- data.trajectory$count
label <- data.trajectory$label

# cell trajectory inference by Slingshot and visualization
# the trajectory plot is saved in the current working directory
FindTrajectory(count, label, title='Contaminated Data')
```

2. **Perform and visualize cell trajectory inference on the clean dataset**

   We use Slingshot to infer the cell trajectories on the dataset `sim_trajectory` after removing all 20% doublets (clean dataset). The following R code shows a two-dimensional visualization of the inference result with two correct cell trajectories ([Figure 5A]).

```
# remove all doublets
count.clean <- count[,which(label==0)]
label.clean <- label[which(label==0)]

# cell trajectory inference by Slingshot and visualization
# the trajectory plot is saved in the current working directory
FindTrajectory(count.clean, label.clean, title='Clean Data')
```

3. **Perform and visualize cell trajectory inference on the post-doublet-detection datasets**

   We first perform doublet detection on the dataset `sim_trajectory` to obtain doublet scores. Then for each method, we remove the top 20% droplets that receive the highest doublet scores. Finally, we infer and visualize cell trajectories on each post-doublet-removal dataset to examine if the corresponding doublet-detection method removes the spurious cell branches formed by doublets ([Figure 5A]).

```
score.list.trajectory <- FindScores(count, methods)
doublet.list.trajectory <- FindDoublets(score.list.trajectory, rate = .2)
data.removal.list.trajectory <-
        RemoveDoublets.Method(count, label, doublet.list.trajectory)

# infer trajectory on each post-doublet-removal dataset
for(method in methods){
    FindTrajectory(data.removal.list.trajectory[[method]]$count,
        data.removal.list.trajectory[[method]]$label, title = method)
}
```

4. **Infer temporally DE genes**

   We first use Slingshot to infer the cell pseudotime on the dataset `sim_temporally_DE` (contaminated dataset). It contains a single cell lineage with 250 temporally DE genes out of 750 genes, mixed with 20% doublets. Second, we use a general additive model (GAM) (Hastie, Tibshirani, and Friedman 2009) to regress each gene's expression levels on the inferred pseudotime. Finally, we calculate the precision, recall, and TNR of the inferred temporally DE genes identified using the Bonferroni-corrected p-value threshold of 0.05. We repeat the same analysis on the clean dataset (without doublets) and each post-doublet-removal dataset.

```
# specify your directory of dataset sim_temporally_DE
data.temp <-
```

```
    readRDS('directory/synthetic_datasets/sim_temporally_DE.rds') # Do not
directly copy this line ("directory" is where "synthetic_datasets.zip" is
downloaded and uncompressed)
count <- data.temp$count
label <- data.temp$label

# ground-truth temporally DE genes
gene.de <- data.temp$gene.de

# calculate precision, recall, and TNR of temporally DE genes for
# contaminated data
de.temp.list <- FindTempDE(count, gene.de)

# calculate doublet scores and remove doublets
score.list.temp <- FindScores(count, methods)
count.clean.temp <- count[,which(label==0)]
label.clean.temp <- label[which(label==0)]
doublet.list.temp <- FindDoublets(score.list.temp, rate=0.2)
data.removal.list.temp <-
                RemoveDoublets.Method(count, label, doublet.list.temp)

# add clean data
data.removal.list.temp[['Clean Data']] <- list(count=count.clean.temp,
                                    label=label.clean.temp)

# calculate precision, recall, and TNR of temporal DE genes for
# post-doublet-removal data
de.temp.result.all <- FindTempDE.All(data.removal.list.temp, gene.de)

# add the result of contaminated data
de.temp.result.all[['Contaminated Data']] <- de.temp.list
```

5. **Visualize the effects of doublet detection on the inference of temporally DE genes**
   We use barplots to compare the results of temporally DE genes identification on the contaminated dataset, the clean dataset, and the post-doublet-removal datasets (Figure 5B). The barplot stacks the results of precision, recall, and TNR for different doublet-detection methods.

```
# transform to data frame for visualization
table.DE.temp <- ListToDataframe(de.temp.result.all, type='barplot')

# draw barplot
# save the plot to a file in the current working directory
Plot_Barplot_temp(table.DE.temp, title='Temporally DE Genes',
                save=T, name = 'temporally DE gene.png', path = getwd())
```

## Performance of doublet-detection methods under distributed computing

**Timing: 8-10 h**

This section illustrates how to use DoubletCollection to evaluate the accuracy of doublet-detection methods under distributed computing. This benchmark simulates the scenario when the large scRNA-seq dataset is beyond the capacity of a single computer so that the dataset must be divided into subsets to be analyzed in parallel.

1. **Calculate detection accuracy under distributed computing**
   First, we randomly split the dataset `pbmc-ch` into two up to ten equal-sized batches. Second, for each batch number, we execute every doublet-detection method on each batch separately and concatenate the resulting doublet scores across batches. Finally, we calculate the distributed AUPRC based on the concatenated doublet scores.

   ```
   # read dataset pbmc-ch
   count <- data.list$count$`pbmc-ch`
   label <- data.list$label$`pbmc-ch`
   label <- ifelse(label == 'doublet', 1, 0)

   # calculate distributed AUPRC for different methods
   auc.list.batch <- FindDistributedAUC.All(count, label, methods,
                       batches=2:10, type='AUPRC')
   ```

2. **Visualize detection accuracy under distributed computing**
   We use line plots to show how the detection accuracy of each method changes as the number of batches increases. The following R code places the batch numbers on the x-axis and connects AUPRC values to show the trend of each method (Figure 5C).

   ```
   # transform the output of FindDistributedAUC.All to a data frame for
   # visualization
   table.batch <- ListToDataframe(auc.list.batch, type='distributed')

   # draw line plots
   # save the plot to the local directory specified by the path parameter
   Plot_Lineplot_Distributed(table.batch, data='pbmc-ch',
           measurement='AUPRC', save=T, name = 'distributed.png',
           path = getwd())
   ```

**Optional**: Users can apply the same pipeline to evaluate the detection accuracy of doublet-detection methods under distributed computing on any other real dataset.

## Computational aspects of doublet-detection methods (optional)

**Timing: 8-10 h**

The benchmark of computational aspects of doublet-detection methods includes but is not limited to efficiency, scalability, stability, and software implementation. First, we can summarize the running time of doublet-detection methods on the 16 real scRNA-seq datasets. The result can be visualized by boxplots similar to Figure 2A to compare the computational efficiency of doublet-detection methods. Second, we can examine how fast each method's running time increases as the number of droplets grows. The result can be visualized by line plots similar to Figure 3A to examine the scalability of doublet-detection methods. Third, we can evaluate how much each method's AUPRC and AUROC values vary across subsets of droplets and genes. The result can be visualized by violin plots to compare the statistical stability of doublet-detection methods. Finally, we can qualitatively evaluate the software implementation of doublet-detection methods from the aspects of user-friendliness, software quality, and active maintenance. The complete visualization details are available in (N. M. Xi and Li 2020).

## Expected Outcomes

The major outcomes of this protocol are the measures of doublet-detection accuracy and the results of downstream analysis, including AUPRC, AUROC, precision, recall, TNR, number of cell clusters, cell trajectories, DE genes, and their visualizations. These outcomes are in the intermediate outputs of the R code shown in previous sections. The visualizations are shown in Figures 2 to 5. More visualizations, tables, and interpretations are available in (N. M. Xi and Li 2020).

Another important result in this protocol is the benchmark of a new method scDblFinder, which was not included in the previous benchmark study (N. M. Xi and Li 2020). On the 16 real RNA-seq datasets, scDblFinder achieves the highest mean AUPRC and AUROC values, and it is also the top method in terms of precision, recall, and TNR under the 10% identification rate. On the synthetic RNA-seq datasets, scDblFinder exhibits similar performance trends to those of other doublet-detection methods under various experimental settings and biological conditions, and it is also a near-top method in terms of AUPRC. In particular, scDblFinder is able to consistently improve downstream analyses, including DE gene, cell clustering, and cell trajectory inference. Similar to other doublet-detection methods, scDblFinder has decreased detection accuracy as the number of batches increases under distributed computing. scDblFinder is also one of the fastest doublet-detection methods (the comparison of running time is not shown). Overall, scDblFinder has excellent detection accuracy and high computational efficiency.

## Limitations

The first limitation of this protocol is that the current benchmark results are based on the default hyperparameters of doublet-detection methods (Lähnemann et al. 2020; N. M. Xi and Li 2020; N. Xi 2021). Therefore, the benchmark results in this protocol may have underestimated the performance of some doublet-detection methods. With the functionality of hyperparameter tuning provided in the R package DoubletCollection, users can conduct an independent study to explore the optimal hyperparameters of doublet-detection methods.

The second limitation of this protocol is that the doublet annotations in the 16 real scRNA-seq datasets are not completely accurate due to experimental limitations. For example, datasets hm12k and hm6k only labeled the heterotypic doublets formed by a human cell and a mouse cell (Zheng et al. 2017); datasets generated by demuxlet only labeled the doublets formed by cells of two individuals (Kang et al. 2018); many homotypic doublets were unlabeled in real datasets (N. M. Xi and Li 2020). The incompleteness of doublet annotations would have inflated the false negative rates and reduced the precision of computational doublet-detection methods. The synthetic datasets used in this protocol contain ground-truth doublets and thus can partly alleviate this issue.

The third limitation of this protocol is that it mainly focuses on doublet-detection methods that can generate a doublet score for every droplet in the dataset. Among currently available doublet-detection methods, DoubletDecon directly outputs identified doublets without providing doublet scores. To fairly compare DoubletDecon with other methods, we suggest users to first execute it on every dataset and record its number of identified doublets; then users can threshold the doublet scores of the other methods so that every method identifies the same number of doublets as DoubletDecon does; finally, users can calculate the precision, recall, and TNR based on the doublets identified by each method from every dataset. A detailed comparison between DoubletDecon and other methods has been discussed by (N. M. Xi and Li 2020). Guidance for executing DoubletDecon is available at (DePasquale et al. 2020).

## Troubleshooting

Users can report any running issues of DoubletCollection to https://github.com/xnnba1984/DoubletCollection/issues. Below we list some common problems we encountered in our testing environment and their solutions.

### Problem 1:
Method Scrublet or DoubletDetection fails to be installed even with a Python 3 environment installed in the system.

### Potential Solution:
This problem typically happens in the Windows system with error information "Microsoft Visual C++ 14.0 is required". To solve this problem, users can download and install the latest version of Visual Studio Build Tools at https://visualstudio.microsoft.com/downloads/ under the menu "Tools for Visual Studio 2019 -> Build Tools for Visual Studio 2019". If Scrublet and DoubletDetection still cannot be installed, users can execute the following two commands in the system shell:

```
pip install scrublet
pip install doubletdetection
```

These commands will manually install Scrublet and DoubletDetection in the Python environment. Then DoubletCollection will be able to call these two methods.

### Problem 2:
The defunction of the doubletCells method.

### Potential Solution:
The doubletCells method is part of the scran package. In the latest version of scran (1.20.1), the scran authors abandoned doubletCells and recommended using scDblFinder method instead. Therefore, if users have already updated their scran packages to version 1.20.1 before installing DoubletCollection, they will not obtain output from doubletCells. If the version of scran is 1.18.7 or older, the doubletCells method will provide a normal output.

## Problem 3:

Some methods fail to generate doublet scores on 16 real scRNA-seq datasets.

## Potential Solution:

This problem is caused by memory shortage when executing certain methods on large-scale datasets. For example, we observed such issues for DoubletFinder on a laptop with 16GB memory. However, using the same code and data, the issue disappears on a server with 256GB memory. To successfully replicate the result in this protocol, we suggest users execute DoubletCollection on a computer with 64GB or more memory. If users perform doublet detection on smaller datasets, then the memory size requirement is less. For example, we select pbmc-1A-dm, pbmc-1B-dm, pbmc-1C-dm, and J293t-dm in one of our preliminary tests. The memory requirement for running on them is much less and thus executable on a standard laptop.

## Problem 4:

The `ReadData` function cannot read scRNA-seq datasets into the R environment.

## Potential Solution:

The `ReadData` function is designed to read all `rds` files under the user-specified directory. Therefore, users need to save all `rds` files in the directory indicated by the `path` parameter of `ReadData`. Users can also use the generic R function `readRDS` to read the single `rds` file.

## Problem 5:

The installation of DoubletCollection will automatically install eight doublet-detection methods and multiple visualization packages. In certain versions of R and Python environments, there are errors caused by those dependent packages. Some examples include:

    a.  Package A XXX version was found, but >YYY is required by YYY
    b.  Error in dyn.load(file, DLLpath = DLLpath, ...): unable to load shared object XXX
    c.  Invalide class XXX object: undefined class for slot YYY

## Potential Solution:

Those errors are caused by the conflicts between the eight doublet-detection methods' packages and other already-installed packages in the environments. Key dependent packages include Seurat, SingleCellExperiment, scran, SummarizedExperiment, BioGenerics, XGBoost, and many others. To address these issues, we have the following suggestions for users.

    1.  Skip the step of updating current dependent packages to their newer versions.

2. Search the exact error information for a solution, as many errors are commonly caused by popular dependents packages. For example, error b is caused by the failure to install Seurat on macOS.

3. Reinstall a clean R and Python (Anaconda) environment with only default packages. Before R installation, remove all files in the "library" folder under the previous R directory. This will solve all the problems caused by dependent packages.

## Problem 6:

The installation gives an error "GNU MP not found, or not 4.1.4 or up."

## Potential Solution:

This error happens in certain versions of Ubuntu systems. It is caused by the lack of libgmp3 library in the Ubuntu system. To solve this issue, users can execute the following command in the system shell:

```
sudo apt-get install libgmp3-dev
```

and reinstall DoubletCollection.

## Problem 7:

No non-system installation of Python could be found.

## Potential Solution:

This message may pop up when the first time FindScores or FindScores.All is called and there is no Python environment installed. Users can input "Y" after the information "Would you like to download and install Miniconda?". If such information does not show up, users can download and install Anaconda at https://www.anaconda.com/products/individual.

## Problem 8:

The DoubletCollection cannot be updated to the latest version.

## Potential Solution:

To install the latest version of DoubletCollection, first remove DoubletCollection from the current R namespace

```
devtools::unload("DoubletCollection")
```

Then delete the old version

```
remove.packages("DoubletCollection")
```

Finally, install the latest version of DoubletCollection from Github

```
devtools::install_github("xnnba1984/DoubletCollection")
```

 

Users can also implement the previous operations in RStudio by (1) restarting the R session; (2) deleting the old version of DoubletCollection in the "Packages" panel.

## Resource Availability

**Lead Contact**
Further information and requests for resources and reagents should be directed to and will be fulfilled by the Lead Contact, Jingyi Jessica Li (jli@stat.ucla.edu).

**Materials Availability**
This study did not generate unique reagents.

**Data and Code Availability**
The datasets used in this study are available at Zenodo
https://zenodo.org/record/4562782#.YI2lhWf0mbg.

The published article includes part of the R code generated during this study. The source code of package DoubletCollection is available at GitHub
https://github.com/xnnba1984/DoubletCollection.

## Acknowledgments

We thank Dr. Bo Li at the University of Texas Southwestern Medical Center (https://www.lilab-utsw.org/research) for bringing our attention to the doublet-detection problem. We also appreciate the comments and feedback from our group members in the Junction of Statistics and Biology at UCLA (http://jsb.ucla.edu). This work was supported by NIH/NIGMS R01GM120507, NSF DBI-1846216, Sloan Research Fellowship, Johnson & Johnson WiSTEM2D Award, and UCLA DGSOM W. M. Keck Foundation Junior Faculty Award.
MSC 2010 subject classifications: 62H20

## Author Contributions

N.M.X. and J.J.L. conceived the idea. N.M.X. performed the bioinformatics analysis and wrote the paper. J.J.L. thoroughly reviewed and edited the paper.

## Declaration of Interests

The authors declare no competing interests.

## Figures and Legends

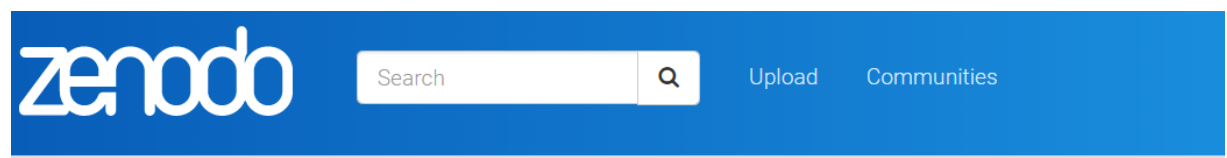

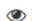

**Figure 1. The Zenodo repository for downloading real and synthetic scRNA-seq datasets used in this protocol.**

**A**

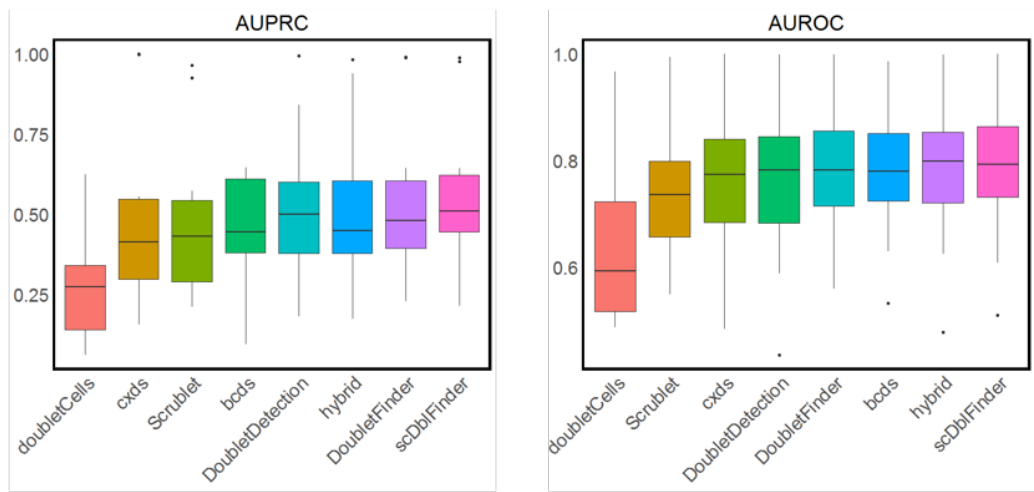

**B**

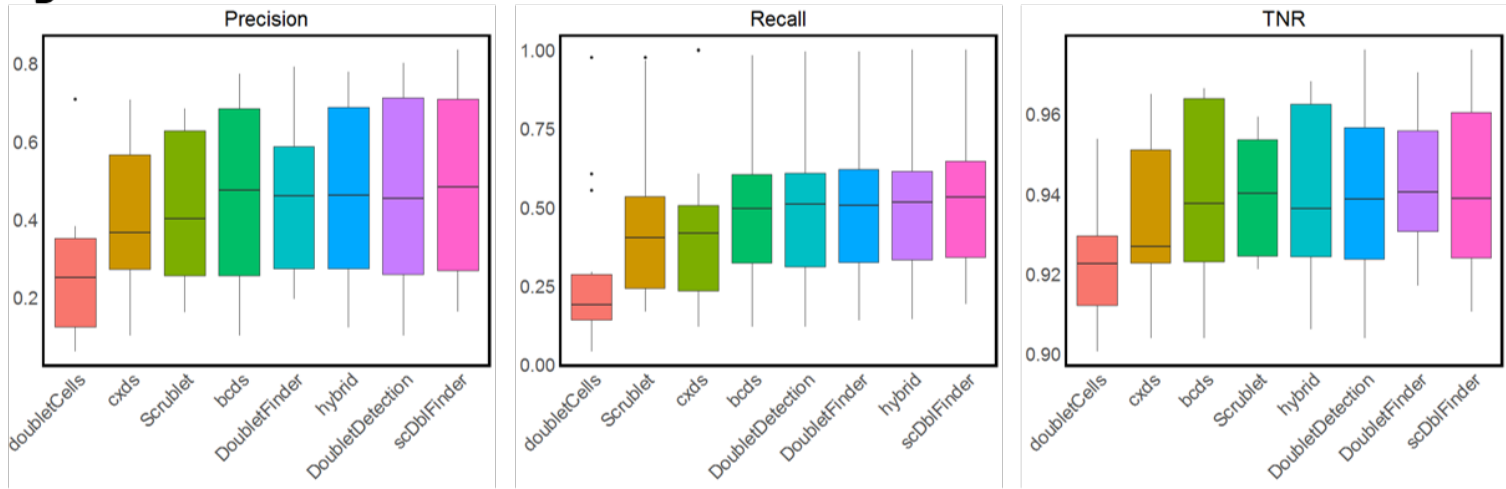

**Figure 2. Evaluation of Doublet-Detection Methods Using 16 Real scRNA-Seq Datasets.** (A) AUPRC and AUROC values of each method applied to 16 datasets. (B) Precision, recall, and TNR values of each method under the 10% identification rate. Methods are ordered by their average performance measurement across 16 datasets (low to high).

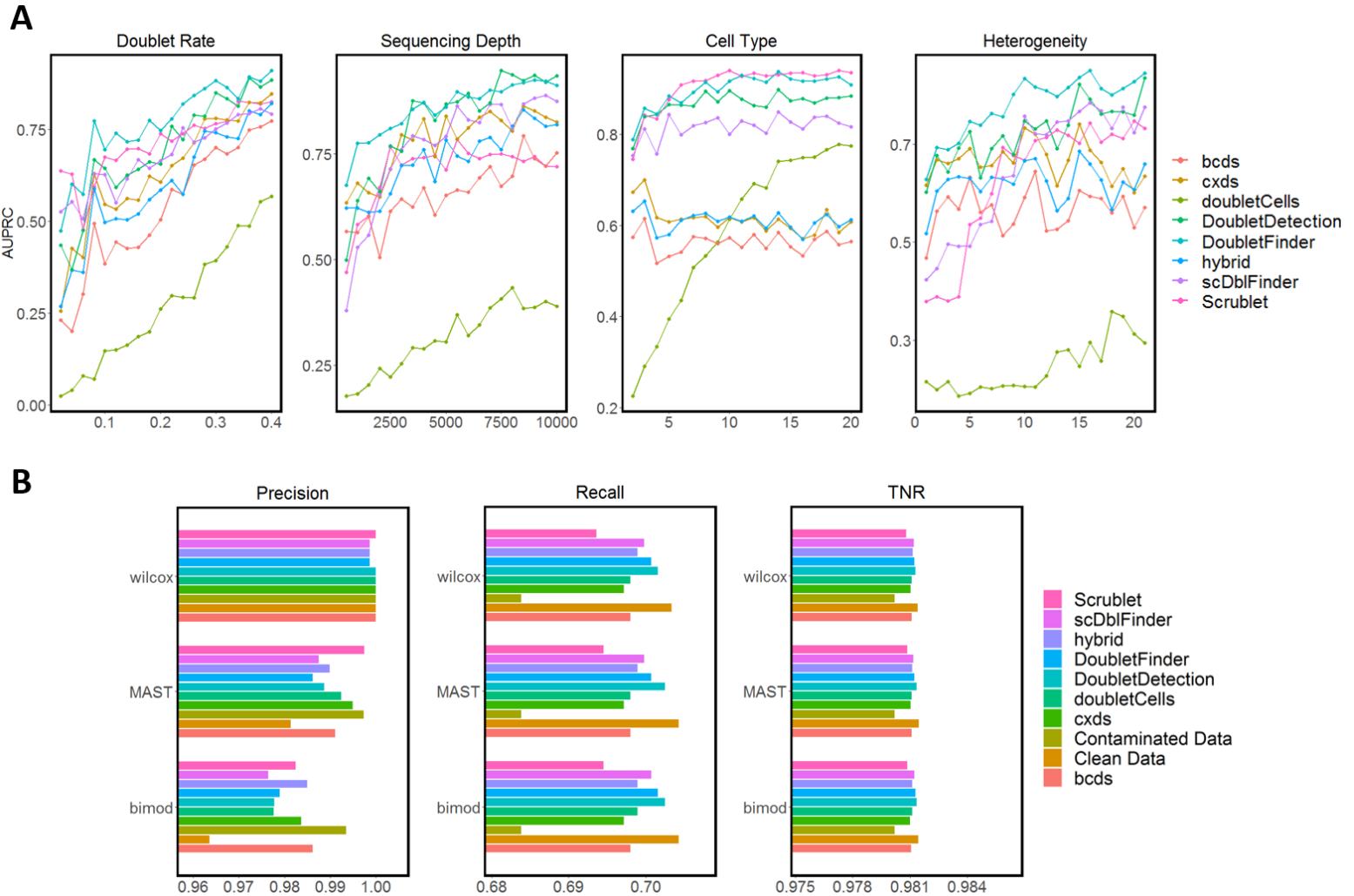

**Figure 3. Evaluation of Doublet-Detection Methods Using Four Simulation Studies, and the Effects of Doublet Detection on DE Gene Analysis.** (A) AUPRC of each method in four simulation settings: varying doublet rates (from 2% to 40% with a step size of 2%), varying sequencing depths (from 500 to 10,000 UMI counts per cell, with a step size of 500 counts), varying numbers of cell types (from 2 to 20 with a step size of 1), and 20 heterogeneity levels, which specify the extent to which genes are differentiated between two cell types. (B) Precision, recall, and TNR by each of three DE methods: Wilcoxon rank-sum test (wilcox), MAST, and likelihood-ratio test (bimod) after each doublet-detection method is applied to a simulated dataset; for negative and positive controls, we included the DE accuracies on the contaminated data with 40% doublets and the clean data without doublets.

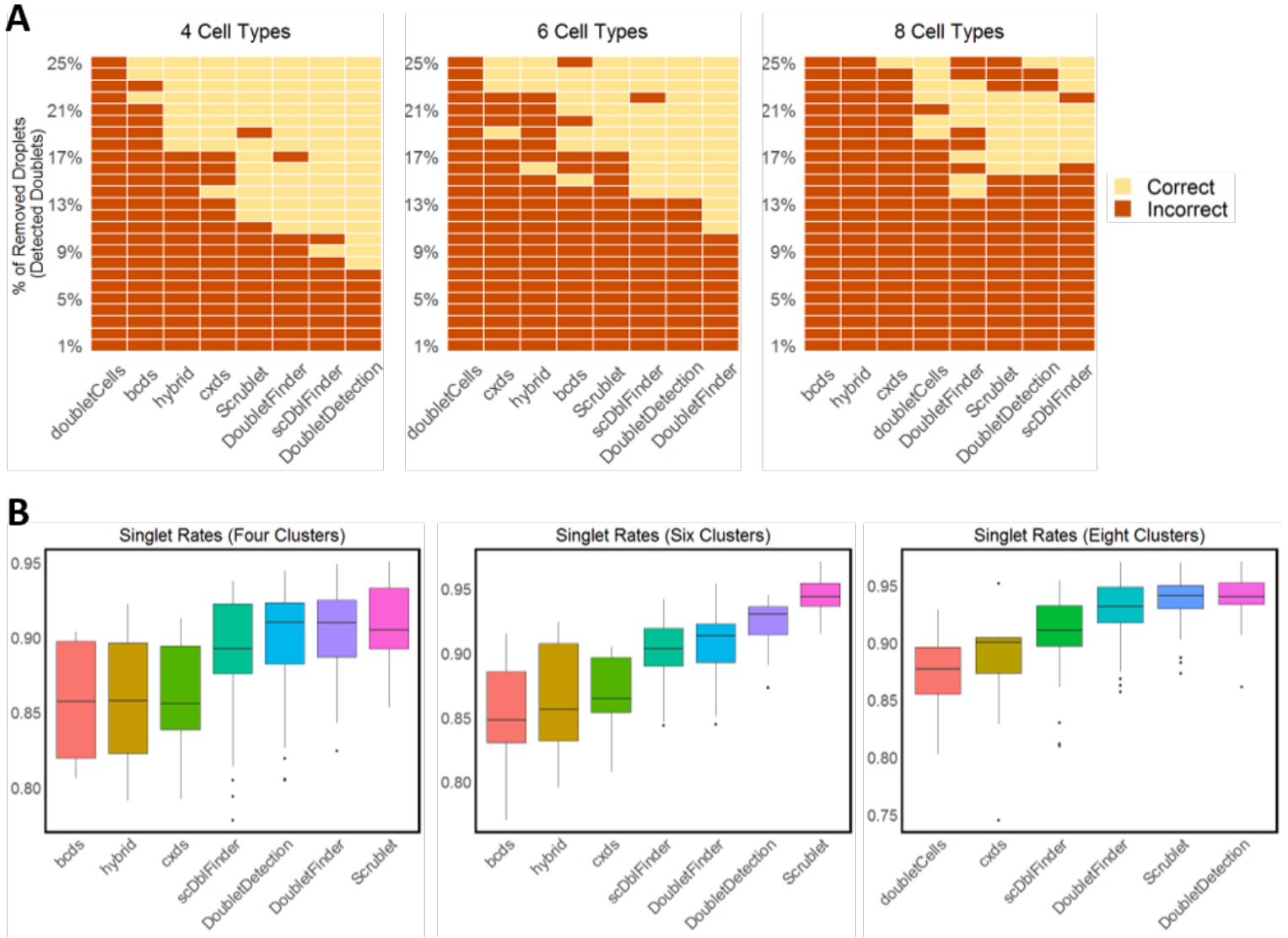

**Figure 4. The Effects of Doublet Detection on Cell Clustering.** (A) Cell clustering results by the Louvain algorithm after each doublet-detection method is applied to remove a varying percentage of droplets as the identified doublets (y-axis, from 1% to 25% with a step size of 1%); the true numbers of cell clusters are four, six, and eight under three simulation settings, each containing 20% true doublets; the yellow color indicates that the correct number of clusters was identified, while the red color indicates otherwise. (B) Under the same three simulation settings as in (A), the distributions of the singlet proportions are shown after doublet removal by each method, if the remaining droplets lead to the correct number of cell clusters in (A); some methods are not shown because they do not lead to the correct number of cell clusters in (A). Methods are ordered by their average performance measurement across 16 datasets (low to high).

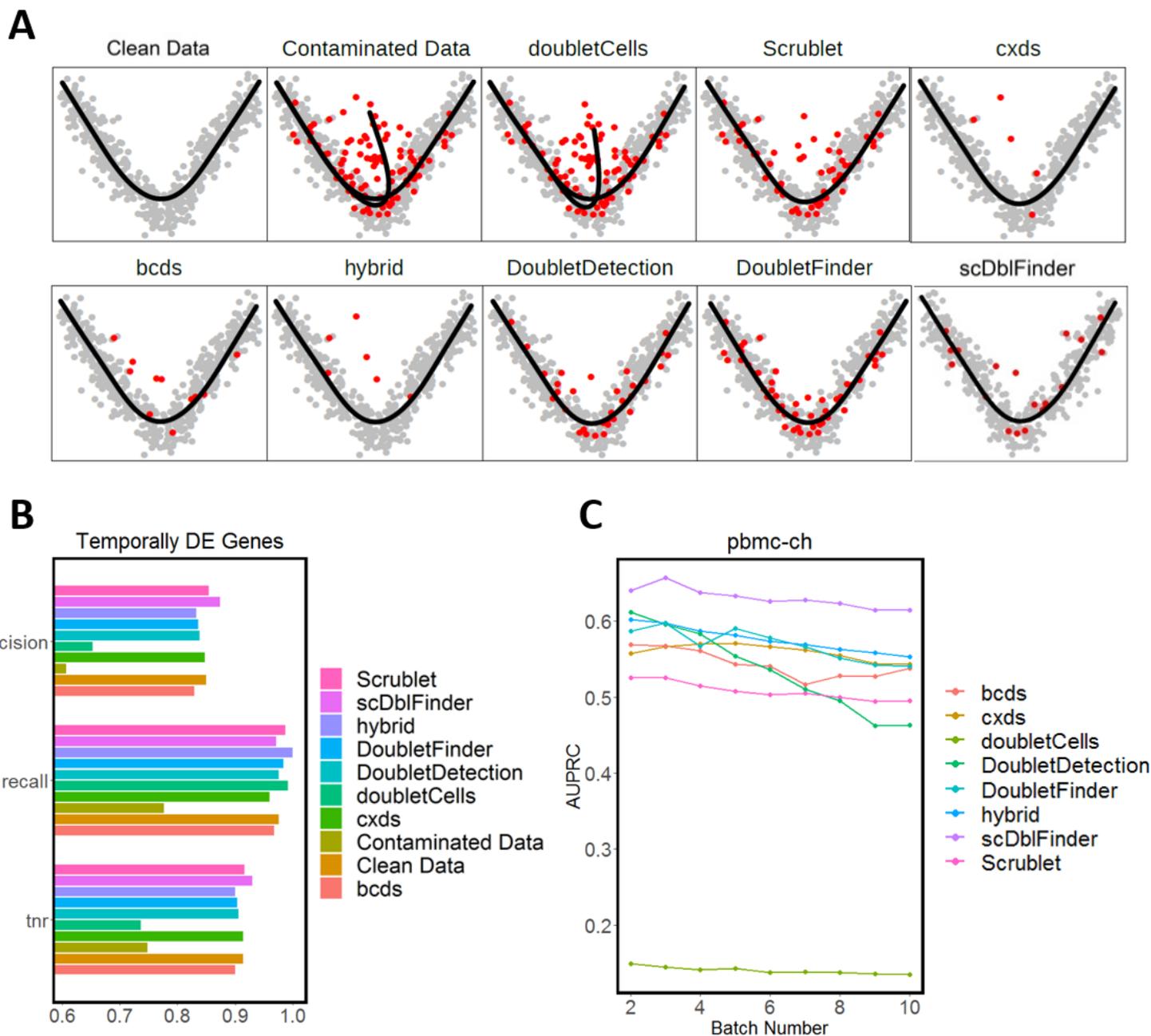

**Figure 5. Effects of Doublet Detection on Cell Trajectory Inference and the detection accuracy under distributed computing.** (A) Cell trajectories constructed by Slingshot. (B) Precision, recall, and TNR of temporally DE genes inferred by the GAM. Both (A) and (B) are performed on contaminated, clean, and post-doublet-detection datasets. (C) AUPRC of each doublet detection method on the real dataset `pbmc-ch` under distributed computing.

**A**

**B**

**Figure 6. The illustration of using the function ReadDate.** (A) The local directory where 16 read datasets are located. (B) The R list with 16 real datasets after executing the function ReadData.